\documentclass[12pt]{article}
\usepackage{amsfonts,amsmath,amssymb}
\usepackage{scalerel}[2016/12/29]
\usepackage{graphicx}
\usepackage{tikz}
\usepackage{braket}

\usetikzlibrary{arrows}
\usetikzlibrary{decorations.markings}

\numberwithin{equation}{section}

\def\be{\begin{equation}}
\def\ee{\end{equation}}
\def\bea{\begin{eqnarray}}
\def\eea{\end{eqnarray}}

\def\bb{\hskip -0.5mm}

\def \bz{{\bar z}}
\def \bx{{\bar x}}

\def \hx{{\hat x}}
\def \bt{{\bar t}}
\def \hht{{\hat t}}
\def \ct{{\cos\hskip -0.3mm\theta\hskip 0.6mm}}
\def \cts{{\cos^2\hskip -0.6mm\theta\hskip 0.6mm}}
\def \st{{\sin\hskip -0.3mm\theta\hskip 0.6mm}}
\def \tt{{\tan\hskip -0.3mm\theta\hskip 0.6mm}}
\def \cp{{\cos\hskip -0.3mm\phi\hskip 0.6mm}}

\usepackage{amsmath}
\usepackage{epsfig,graphicx}
\usepackage{graphicx}
\usepackage{tikz}

\def\be{\begin{equation}}
\def\ee{\end{equation}}
\def\bea{\begin{eqnarray}}
\def\eea{\end{eqnarray}}

\begin{document}

\thispagestyle{plain}

\title{\bf\large Superluminal anisotropic propagation and wavefront splitting on tilted and boosted braneworlds}

\author{Alexios P. Polychronakos}

\date{ }

\maketitle

%

{\em
{\centerline{Physics Department, the City College of New York, NY 10031, USA}}
\vskip .2 cm \centerline{and}
\vskip .2 cm
{\centerline{The Graduate Center, CUNY, New York, NY 10016, USA}
\vskip .3 cm
\centerline{\it apolychronakos@ccny.cuny.edu}
\centerline{\it apolychronakos@gc.cuny.edu}
}}

\maketitle

\begin{abstract}

Braneworlds winding and spinning around an extra compact dimension manifest superluminal
propagation for fields penetrating the bulk. This propagation is either irreducibly anisotropic or one that
becomes isotropic in a special frame, depending on the brane's motion and orientation in the bulk.
For a class of boosted observers on the brane the wavefront of such fields will
split into two hyperboloid components, one propagating forward and the other backward in time at
superluminal speeds. In spite of these effects, there is no violation of causality and no tachyons.
Detection of astrophysical anisotropic superluminal propagation velocities would offer a sign for the existence of
extra compactified dimensions.

\end{abstract}

\vskip 1cm

\vfill
\eject

\tableofcontents

\section{Introduction}

Braneworlds on compactified spacetimes can have some counterintuitive properties due to the
possibility of fields propagating over the bulk \cite{GrPa}. Recently, Greene {\it et al.} \cite{Kab} considered
the fun problem of light propagation in a flat $M_4$ braneworld embedded in the bulk spacetime
$M_4 \times S^1$ \cite{Dim} and
pointed out that, if the braneworld is boosted, light would travel superluminally on the brane. Superluminal
transmission is usually associated with acausal effects, and yet there is no violation of causality on the brane.

The essense of what happens is captured by the ``scissors effect": the intersection of the blades of a closing
scissors can move faster that the speed of light as the blades become parallel. This is rightfully dismissed as
a non-physical effect, since the intersection point carries no energy or physical information. However, in the
case at hand, the propagating wavefront ``scissoring'' the brane was originally emitted by the brane, thus
carrying energy and information. In effect, the compact geometry of the bulk promotes the scissors effect into
a (continuous) wormhole effect, resulting in superluminal propagation.

In this work we extend this effect to the most general situation in which branes are both boosted
and tilted with respect to the compact dimension; that is, braneworlds that wind around and spin around
the extra compact dimension (kind of like a barber's pole).
The qualitatively new feature that emerges in this setup is that
braneworlds fall generically into two distinct classes: ``boostlike'' ones, where light propagation is isotropic
in an appropriate frame, and ``tiltlike'' ones with an irreducible light propagation anisotropy.
The separator between the two classes is a special case that, in an appropriate lightcone frame,
exhibits isotropy and propagation at the speed of light as if it were untilted and unboosted. For
a class of Lorentz frames on such braneworlds the light wavefront emitted by a point source splits into two
hyperboloidal components, one propagating forward and the other backward in coordinate time at
superluminal speeds.
In all situations causality is preserved, although all light propagation is superluminal, and there are no tachyons.
The detection of anisotropy and superluminal deviation from the speed of light in cosmological
data or astrophysical observations would offer a sign for the existence of extra dimensions.

\section{Tilted and boosted braneworld}\label{Tilbo}

We will analyze the propagation of bulk lightlike fields on the braneworld starting with a brane in a frame
tilted and boosted
with respect to the compact dimension. Other frames, including one that minimizes
anisotropy, will be studied in the next section.

\subsection{Winding and spinning brane}\label{WinSpin}

Consider a flat spacetime $M_4 \times S^1$ and a flat 4-dimensional braneworld in it that is tilted by an angle
$\theta$ and boosted by a speed $v$ along the compact dimension. We single out the time coordinate $t$,
the compact space coordinate $z$, another space coordinate $x$ along which the braneworld tilts, and the
remaining two untilted braneworld coordinates collectively called $y$.
The speed of light will be set to $1$.

We will generate the tilted and boosted brane by starting from a flat frame
$\bx, \bz, \bt$ in which the periodicity of $S^1$ is expressed by the identification
\be
(\bx, y, \bz, \bt) \sim (\bx, y, \bz +R, \bt)
\ee
with $R$ the length of the compact dimension. An untilted and unboosted brane would lie on the $\bz = 0$
hyperplane in these coordinates. Correspondingly, a tilted and boosted brane would lie on the $z=0$ hyperplane
of an appropriately rotated and boosted frame. Performing a Lorentz transformation consisting of a rotation by an
angle $\theta$ in the $\bx - \bz$ plane and a boost by a speed $v$ in the $\bz - \bt$ plane ($y$ coordinates are not touched), we obtain the brane-fixed coordinates $x,z,t$ as
\bea
x &=& \ct \bx + \st \bz \cr
z &=& \gamma \left(\ct \bz - \st \bx - v \bt\right) \label{TB}\\
t &=& \gamma \left(\bt - \ct v \bz + \st v \bx\right)
\nonumber
\eea
with $\gamma = 1/\sqrt{1-v^2}$ as usual. The position of the brane is at $z =0$, and from (\ref{TB}) we see that
this implies
\be
\bz = \tt \bx + {v \over \ct}\, \bt
\ee
on the brane. This demonstrates that the brane is tilted by an angle $\theta$ with respect to $\bz$ and
moves normal to itself with a speed $v$. Note that the speed of the ``foot'' of the brane on the $\bz$ axis
is $v/\ct$ which could become bigger than $1$, a pure scissors effect. In the brane-fixed coordinates the
periodicity of the bulk becomes
\be
(x,y,z,t) \sim (x+\st R,y,z+\gamma\, \ct R,t -\gamma v\, \ct R)
\ee

\subsection{Light propagation on the bulk}

Consider the emission of a light wave from a point source on the brane, which by assumption
propagates as a spherical wave in the bulk. (Emission of light point particles in the bulk that return and hit the brane can also be considered, but we will focus on waves.) Assuming that the source was at
$x=y=z=t=0$, which is also $\bx=y=\bz=\bt=0$, the periodicity $\bz \sim \bz +R$ implies that in
the covering space of the bulk there is a line of point emitters at $\bz = R n$, $n=0,\pm1,\pm2 \dots$
emitting spherical waves at the same time $\bt =0$. After time $\bt = R/2$ the waves will overlap and,
eventually, their interference will result in the cylindrical wavefront caustic
\be
\bx^2 + y^2 - \bt^2 = 0
\label{cylind}\ee
in a manifestation of the Huygens principle.

An alternative way to see this is to expand the source into Kaluza-Klein modes around the compact
dimension:
\be
\sum_{n=-\infty}^\infty \delta (\bz -nR) = {1\over R} \sum_{m=-\infty}^\infty e^{2\pi i\, m \bz}
\ee
(This is essentially the method of images, as done in \cite{Leon} and repeated in \cite{Kab}.)
The zero mode $m=0$ is massless on the brane and will propagate at the speed of light in the direction normal to $\bz$,
while higher modes are massive and propagate subluminally, so they will eventually be left behind.
For $\bt \gg R$ only the constant mode will survive with a wavefront as in (\ref{cylind}).

\subsection{Light propagation on the brane}

The cylindrical wave (\ref{cylind}) will intersect the brane at $z=0$ and produce a wavefront on the brane.
Its shape can be found by expressing the original coordinates $\bx,\bz,\bt$ in terms of $x,z,t$ by inverting
(\ref{TB}), 
\bea
\bx &=& \ct x -\gamma \st (z+vt) \cr
\bz &=& \st x +\gamma \ct (z + vt)\cr
\bt &=& \gamma (t +v z)
\nonumber
\eea
setting $z=0$, and substituting in (\ref{cylind}). We obtain
\be
\left(\ct x - \gamma v\, \st t \right)^2 +y^2 - \gamma^2 t^2 = 0
\label{wav}\ee
We see that this wavefront is anisotropic and has a drift. Its spatial sections are ellipsoids with axes
of ratio $\ct$, and their center drifts in the $x$ direction with a speed $v_d = \gamma v \, \tt$.
The speed of propagation $c_\phi$ in a direction at an angle $\phi$ from the $x$ axis is
\be
c_\phi = {\gamma v\, \ct \st \cp+ \sqrt{\gamma^2 \cos^2\hskip -0.4mm\theta +
\sin^2\hskip -0.4mm\theta \sin^2\hskip -0.4mm\phi} 
\over 1-\sin^2\hskip -0.4mm\theta \cos^2\hskip -0.4mm\phi}
\label{cdefault}\ee
Specifically, the propagation speed in the $+x$, $-x$, and $y$ directions is
\be
c_{\pm} = \gamma {1\pm v \,\st \over \ct} ~,~~~ c_y = \gamma \sqrt{1-v^2 \sin^2\hskip -0.3mm\theta}
\label{propaxy}\ee
We remark that all the above speeds are greater than $1$.
If $|v| \le |\st\bb|$, then $c_\phi$ will become equal to the
speed of light in the unique direction $\phi_o$
\be
\cp_o = -\gamma v \cot\hskip -0.3mm\theta ~~~\Rightarrow~~~ c_{\phi_o} = 1
\label{po}\ee
For $v= \st$, $c_{-} =1$, and for $v=-\st$, $c_{+} =1$. For all other angles, and for all angles when
$|v| > |\st\bb|$, propagation speeds are superluminal.

\section{Boosted frames on the brane}

The form of the light wavefront and the corresponding propagation speeds depend on the frame of observers on the brane. The frame of the previous section was the ``default'' frame in which observers' bulk velocity is
normal to the brane. Other observers, boosted on the brane, can be considered. We will examine the general
case of such observers in this section. Moreover, we will identify a special frame in which the anisotropy of
light propagation is either eliminated or minimized. This is useful for phenomenological considerations,
but also for revealing the two qualitatively distinct cases of brane tilting and boosting.

\subsection{A driftless frame}\label{driftless}

We can eliminate the anisotropy and/or the drift in the wave propagation by performing an additional
boost in the $x - t$ plane into new coordinates ${\hat x}, {\hat t}$, which leaves the brane position and orientation unchanched at $z=0$.
We set
\bea
x &=& g \left( \hx - u \hht \hskip 0.04cm\right) ~~~,~~~g = {1\over \sqrt{1-u^2}}\cr
t &=& g \left( \hht - u \hx \right)
\label{boostu}\eea
and adjust the boost $u$ to eliminate the cross-term ${\hat x} \cdot {\hat t}$ in (\ref{wav}). We need to consider
two separate cases:

\subsubsection{Tiltlike anisotropic case: $|v| < |\st\bb|$}

Performing the transformation (\ref{boostu}) on (\ref{wav}) and setting the coefficient of the term
$\hx \cdot \hht$ to zero we obtain
\be
u = \gamma v \cot \theta ~,~~~ g = {\st \over \gamma \sqrt{\sin^2\bb\theta - v^2}}
\ee
and (\ref{wav}) becomes
\be
\gamma^2 \cos^2\bb\theta \,\, \hx^2 + y^2 = \hht^2
\label{frontilt}\ee
We see that the drift is eliminated, but wave propagation remains anisotropic. The speed of propagation at
angle $\phi$ with respect to the $\hx$ axis is
\be
c_\phi = {1\over \sqrt{\sin^2\hskip -0.4mm\phi + \gamma^2 \cos^2\hskip -0.4mm\phi \cos^2\hskip -0.4mm\theta}}
\label{cxy}\ee
The two extreme propagation speeds are
\be
c_\hx = {1 \over \gamma\, \ct} >1 ~,~~~ c_y = 1
\ee
We also note that no additional Lorenz boost can make the wavefront isotropic, even at the price of
introducting a drift. The periodicity of bulk space in this frame is
\be
(\hx,y,z,\hht) \sim (\hx,y,z,\hht) + \left(\gamma \sqrt{\sin^2\bb\theta - v^2},0, \gamma\, \ct , 0\right) R
\label{at}\ee

\subsubsection{Boostlike isotropic case: $|v|> |\st\bb|$}

In this case the elimination of $\hx \cdot \hht$ is achieved with the boost
\be
u = {1 \over \gamma v} \tt ~,~~~ g = {v \,\ct \over \sqrt{v^2 - \sin^2 \theta}}
\ee
and the wavefront becomes
\be
\hx^2 + y^2 = \gamma^2 \cos^2\bb\theta \, \hht^2
\label{fronboost}\ee
We note that this is a driftless, isotropic wave with a uniform propagation speed
\be
c = \gamma \,\ct >1
\label{cc}\ee
The periodicity of bulk space in this case is
\be
(\hx,y,z,\hht) \sim (\hx,y,z,\hht) + \left(0,0, \gamma \,\ct , -\gamma \sqrt{\sin^2\bb\theta - v^2}\right) R
\label{ab}\ee

The limiting case $|v| = |\st\bb|$ is interesting. Elimination of cross-terms in this case requires a boost
$u=1$, apparently a singular limit. Nevertheless, the brane spacetime remains regular in this limit,
light propagates at a speed $c=1$, and the periodicity of bulk space is simply $z \sim z +R$, just as
in the untilted unboosted braneworld.

It should be clear that the above three cases exhaust all possibilities for the light propagation dynamics
on a braneworld embedded in bulk space $M_4 \times S^1$. Adopting coordinates in the bulk such that
the brane is at position $z=0$, as we can always do, the compactification of the bulk space can
be expressed as
\be
(x,y,z,t) \sim (x,y,z,t) + (a_x, a_y, a_z , a_t)
\ee
The five-vector $\vec a$ is necessarily spacelike, otherwise the bulk spacetime would have closed timelike
curves guaranteeing the violation of causality. With a rotation within the brane we can set $a_y = 0$.
Then a boost along $x$ can eliminate one of the other components of $\vec a$: if $(a_x , a_t )$ is
spacelike we can eliminate $a_t$, bringing $\vec a$ to the form (\ref{at}) of a tiltlike braneworld;
if $(a_x ,a_t )$ is timelike we can eliminate $a_x$ and bring it to the form (\ref{ab}) of a boostlike braneworld.
The special intermediate case corresponds to a timelike $(a_x , a_t)$. An infinite boost can drive it to
zero, leaving $a_z$ as the only nonzero periodicity variable.

\subsection{General boosted frames and split wavefronts}

For a general boost (\ref{boostu}), with $u$ unrelated to $v$ and $\theta$, the lightfront (\ref{wav}) becomes
further deformed, with some interesting qualitative features. To uncover them, we will start from the special
driftless frame $(\hx, \hht)$ of section {\bf\ref{driftless}} and perform a boost. This simplifies the calculations substantially
over working with the default frame of section {\bf\ref{WinSpin}}, the only price
being that we need to consider separately the tiltlike and boostlike cases.

\subsubsection{Boosted frame on tiltlike brane}

The boost of a boosted frame would be, in general, along an arbitrary direction in the $(x,y)$ space on the brane.
However, any such boost can be obtained as a boost along the $y$ directions followed by a boost along the
$x$ direction. Since the wavefront (\ref{frontilt}) is invariant under boosts in $y$, it is enough to consider
boosts along the $x$ direction of the form
\bea
x &=& g \left( \hx + u \hht \hskip 0.04cm\right) ~~~,~~~g = {1\over \sqrt{1-u^2}}\cr
t &=& g \left( \hht + u \hx \right) \label{hat}
\eea
where we used the notation $x,t$ for the general boosted coordinates (not to be confused with the ``default''
frame of section {\bf\ref{Tilbo}}). The wavefront (\ref{frontilt}) in the boosted frame becomes
\be
\gamma^2 \cts (x+u t)^2 +{y^2\over g^2} - (t+ux)^2 =0
\ee
This is a quadratic form that can be brought to the canonical form
\be
(\gamma^2 \cts - u^2 ) \, x^2 +2u(\gamma^2 \cts -1)\, xt + (\gamma^2 \cts u^2 -1) \, t^2 + (1-u^2)\, y^2
= 0
 \label{split}\ee
We see that the form of the wavefront depends crucially on the sign of the coefficient of $x^2$
$\gamma^2 \cts - u^2$ (note that $\gamma |\ct\bb|<1$ in the tiltlike case). For $|u| < \gamma |\ct\bb|$,
the equal-time wavefronts are elongated ellipsoids with a drift. As $u$ approaches one of
the critical values $\pm \gamma |\ct\bb|$, the ellipsoid elongates further and becomes a paraboloid, and
for $|u| > \gamma |\ct\bb|$ the wavefront splits into the two branches of a hyperboloid.

However, only one of the branches of this hyperboloid exists for $t>0$: the wavefront was emitted
from a point source at time $\bt =0$ in (\ref{cylind}) and propagates for $\bt>0$. The condition $\bt >0$
translates to the condition
\be
( 1 - \gamma v u \cot \theta ) \, t + (u - \gamma v \cot \theta) \, x >0
\label{tbpostilt}\ee
in the boosted frame of the tiltlike brane, and only one of the branches of the hyperboloid satisfies this
condition. To see this, note that the intercepts $x_\pm$ of the wavefront on the $x$ axis ($y=0$) are
\be
x_\pm = - {\gamma \ct u \pm 1 \over \gamma \ct \pm u} \, t
\ee
and on the intercepts the $\bt >0$ condition (\ref{tbpostilt}) becomes
\be
\gamma (1-u^2 ) {1 \pm v \cot \theta \over \gamma \pm {u/ \ct}} >0
\label{tpostil}\ee
Due to the tiltlike condition $|v| < |\st\bb|$, the numerator is always positive. 
For $t>0$ and $|u| < \gamma |\ct\bb|$, the condition (\ref{tpostil}) is satisfied for both $x_\pm$. In this case the wavefront is
an ellipsoid with $x_\pm$ its two interecepts on $x$, and the full ellipsoid wavefront corresponds to
$\bt >0$. However, for $|u| >\gamma |\ct\bb|$, only $x_+$ satisfies the condition for $\ct >0$, and only
$x_-$ satisfies it for $\ct <0$. In this case the wavefront is a hyperboloid, and only the branch with the
corresponding intercept survives.

The situation differs for $t<0$: for $|u| < \gamma |\ct\bb|$ none of $x_\pm$ satisfies the condition
$\bt >0$ and there is no wavefront. However,  for $|u| >\gamma |\ct\bb|$, $x_-$ satisfies the condition
if $\ct >0$ and $x_+$ does if $\ct <0$. So the wavefront for $t<0$ is obtained by setting $x\to -x,\,t\to -t$
in the branch of the hyperboloid rejected for $t>0$.

We conclude that for $u> \gamma |\ct\bb|$ the wavefront splits into two
hyperboloids, one propagating forward and the other backward in coordinate time. This is a peculiar
effect, but it does not lead to causality violation, as will be demonstrated in the next section.

The propagation speed $c_\phi$ at an angle $\phi$ with respect to the $x$ axis can be calculated from
(\ref{split}) by putting $x=\cos\phi\, c_\phi\, t$ and $y= \sin\phi \,c_\phi\, t$.
We obtain
\be
c_\phi = {u(1-\gamma^2 \cts )\cos\phi +\sqrt{(1-u^2)[\gamma^2 \cts (\cos^2\phi - u^2) + \sin^2\phi]}
\over \gamma^2 \cts \cos^2\phi +\sin^2\phi - u^2}
\label{ctilt}\ee
For $\phi = 0,\pi$, $c_\phi$  reduces to $x_\pm / t$. We note that in the case $|u| > \gamma |\ct\bb|$
there is a critical angle $\phi_c$ for
\be
\tan \phi_c = -{\rm sgn} (u)\, g \sqrt{u^2 - \gamma^2 \cts}
\ee
beyond which light propagation is not possible for $t>0$: only directions $\phi < \phi_c$ (for $u>0$;
otherwise $\phi >\phi_c$) are accessible to the
wavefront. For $\phi \to \phi_c$ the propagation speed diverges, and $\phi > \phi_c$ corresponds
to the part of the wavefront propagating backward in time. We also note that, for all $u$,
there is a unique direction $\phi_o$
\be
\cos \phi_o = - u
\ee
for which $c_{\phi_o} =1$. For all other angles $c_\phi >1$.

\subsubsection{Boosted frame on boostlike brane}

For a boostlike brane the wavefront propagation (\ref{fronboost}) is isotropic and all boosts are equivalent.
We can again, then, choose a boost along the $x$ direction. The boosted wavefront becomes
\be
(x+ut)^2 +{y^2 \over g^2} - \gamma^2 \cts (t+ux)^2 = 0
\ee
which can be brought to the canonical form
\be
(1-\gamma^2 \cts u^2 ) \, x^2 + 2u (1-\gamma^2 \cts ) \, xt + (u^2 - \gamma^2 \cts )\, t^2 +
(1-u^2) \,y^2 = 0
\label{splib}\ee
The situation is qualitatively similar to the tiltlike case, depending on the sign of the coefficient of
$x^2$, $1-\gamma^2 \cts u^2$. For $|u| < \gamma^{-1} |\ct\bb|^{-1}$ lightfronts will be ellipsoids, but for
$|u| > \gamma^{-1} |\ct\bb|^{-1}$ they will be hyperboloids splitting
into a branch propagating forward and one propagating backward in coordinate time.
The emission condition $\bt > 0$ in the boostlike case translates to the condition
\be
\left(1-{u \tan \theta \over \gamma v} \right)\, t +\left(u - {\tan \theta \over \gamma v}\right) \, x >0
\label{tbposboos}\ee
The intercepts of the wavefront on the $x$ axis are
\be
x_\pm = - {u \pm \gamma \ct \over 1\pm \gamma u \ct}\, t
\ee
and on the intercepts the $\bt >0$ condition (\ref{tbposboos}) becomes
\be
(1-u^2 )\, {1 \pm \st /v \over 1 \pm \gamma u \ct} >0
\label{tposboos}\ee
Due to the boostlike condition $|v| > |\st |$, the numerator is always positive.
For $t>0$ and $|u| < 1/(\gamma |\ct\bb|)$, the condition (\ref{tposboos}) is satisfied for both $x_\pm$. In this case the wavefront is
an ellipsoid with $x_\pm$ its two interecepts on $x$, and the full ellipsoid wavefront corresponds to
$\bt >0$. However, for $|u| >1/(\gamma |\ct\bb|)$, only $x_+$ satisfies the condition for $\ct >0$, and only
$x_-$ satisfies it for $\ct <0$. In this case the wavefront is a hyperboloid, and only the branch with the
corresponding intercept survives. For $t<0$ and $|u| < 1/(\gamma |\ct\bb|)$ none of $x_\pm$ satisfies the
condition $\bt >0$ and there is no wavefront, while for $|u| >1/(\gamma |\ct\bb|)$, $x_-$ satisfies the condition
if $\ct >0$ and $x_+$ does if $\ct <0$; the rejected branch of the hyperboloid for $t>0$ gives the $t<0$
wavefront upon $x\to -x$.

In conclusion, as in the tiltlike case, the wavefront for boost speeds above the critical value $1/(\gamma |\st |)$
splits into two
hyperboloids, one propagating forward and the other backward in coordinate time, again without leading
to causality violation.

The propagation speed $c_\phi$ at an angle $\phi$ with respect to the $x$ axis can again be calculated from
(\ref{splib}) by putting $x=\cos\phi\, c_\phi\, t$ and $y= \sin\phi \,c_\phi\, t$ and we obtain
\be
c_\phi = {u(\gamma^2 \cts -1)\cos\phi +\sqrt{(1-u^2)[\gamma^2 \cts (1-u^2 \cos^2\phi) -u^2 \sin^2\phi]}
\over 1-\gamma^2 u^2 \cts \cos^2\phi - u^2 \sin^2\phi}
\label{cboost}\ee
Similarly to the tiltlike brane, in the case 
$|u| > \gamma |\ct\bb|$ there is a critical angle $\phi_c$
\be
\tan \phi_c = -{\rm sgn} (u)\, g \sqrt{\gamma^2 u^2 \cts -1}
\ee
beyond which light propagation only happens for $|\phi| < \phi_c$ for $t>0$, and for $|\phi | > \phi_c$ for
$t<0$ (for $u>0$, vice versa otherwise).
However, unlike in the tiltlike case, $c_\phi > 1$ for all directions and there is no angle at which
$c_\phi = 1$.

\section{Photons and causality}

It should be immediately obvious that no causality violation is possible in the previous braneworlds,
regardless of superluminal propagation. As also noted in \cite{Kab}, images of the source situated
at positions ${\vec X} = n {\vec a}$ in the bulk are spacelike-separated, since $\vec a$ is spacelike,
and cannot influence each other's past. Superluminal wave propagation may be disturbing but it
does not violate causality.

\subsection{Dispersion relations}

Causality issues are clarified by considering the dispersion relation of waves propagating at
the speed of light in the bulk. A plave wave in the bulk would have the form
\be
\phi = \exp i(k_x x + k_y y + k_z z - k_t t )~,~~~ k_x^2+k_y^2+k_z^2-k_t^2 =0
\ee
This is valid in any Lorentz frame. For a wave emitted from the brane, as was explained in section {\bf 2.2},
wave propagation will be normal to $\vec a$, so the additional condition ${\vec k} \cdot {\vec a}=0$
holds. Moreover, on the brane $z=0$. Altogether we have
\be
\phi = \exp i(k_x x + k_y y - k_t t )~,~~~ {\vec k}^2 =0 ~,~~ {\vec k} \cdot {\vec a} =0
\ee
The condition ${\vec k} \cdot {\vec a} =0$ can be solved for $k_z$, and substituting in the light-cone
relation ${\vec k}^2 =0$ yields the dispersion relation for the wave on the brane. For the special
drift-free/isotropic frame $\hx, \hht$ the periodicity vector simplifies and we obtain:

a) Tiltlike case:
\be
{k_x ^2 \over \gamma^2 \cos^2 \bb \theta} + k_y^2 - k_t^2 = 0 ~,~~~ \gamma \,\ct <1
\ee

b) Boostlike case:
\be
k_x^2 + k_y^2 - {k_t^2 \over \gamma^2 \cos^2\bb\theta} = 0 ~,~~~ \gamma \,\ct >1
\ee
The group velocities derived from the above dispersion relations reproduce the propagation velocities
(\ref{cxy},\ref{cc}) obtained before. Upon quantization,
they would lead to photons of energy $E = \hbar k_t$ and momentum $(p_z,p_y) = \hbar (k_x,k_y)$
such that (with $p^2 = p_x^2 + p_y^2$)
\bea
&(a)& ~~E^2 - p^2 = \left({1\over \gamma^2 \cos^2 \bb\theta} -1\right) p_x^2 > 0 \cr
&(b)& ~~E^2 - p^2 = (\gamma^2 \cos^2\bb\theta  -1) p^2 >0
\eea
Therefore, although photons propagate superluminally, they are not tachyonic. What in fact happens
is that when a photon is captured by an observer on the braneworld its full energy is released, while
only the momentum along the dimensions of the brane is detected. The photon energy contributed
by the momentum component normal to the brane results in $E > p$, without making the photon massive.

\subsection{Causality}

Superluminal speeds can give the impression of causality violation since a particle exchanged between
two observers at a speed larger than 1 would change direction of propagation in a
boosted frame, therefore appearing as emitted by the receiver and captured by the emitter, for
an apparent reversal of cause and effect (a phenomenon sometimes called ``traveling backward in time''). Yet this is simply a perception of the boosted observer, and there is no real causality violation.
We will demonstrate this
%
by considering photon properties in a general frame as well as communication between observers.
{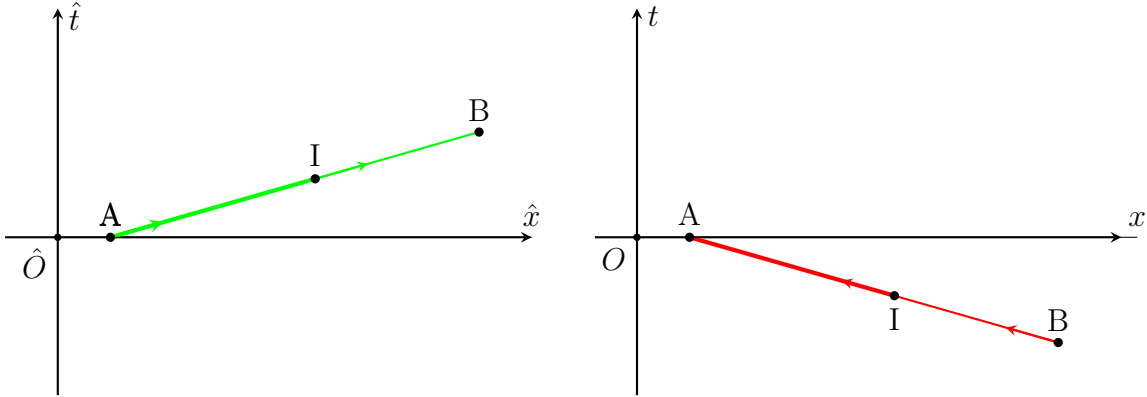
\begin{figure}\vskip-0.5cm
\centering 
{
\begin{tikzpicture}[scale=0.7]


\tikzset{big arrow/.style={decoration={markings,mark=at position 1 with {\arrow[scale=3,#1,>=stealth]{>}}},postaction={decorate},},big arrow/.default=black}

\draw[black,thick,-stealth](-1,0){--++(10,0)};
\draw[black,thick,-stealth](0,-3){--++(0,7.35)};

\draw[green,very thick,-stealth](1,0){--++(1,2/7)};
\draw[green,thick,-stealth](1+7/1.8,2/1.8){--++(1,2/7)};
\draw[ultra thick,green,-](1,0) node[above]{\color{black}{A}}--(1+7/1.8,2/1.8) node[above]{\color{black}{I}};
\draw[black](0,0)--(9,0) node[above]{$\hx$};
\draw[black](0,0)--(0,4.2) node[right]{$\hht$};
\draw[thick,green,-](1,0) node[above]{\color{black}{A}}--(8,2) node[above]{\color{black}{B}};
\fill (1+7/1.8,2/1.8) circle (2.5pt);
\fill (1,0) circle (2.5pt);
\fill (8,2) circle (2.5pt);
\fill (0,0) circle (2pt) node[below left]{$\hat O$};

\draw[black,thick,-stealth](10.2,0){--++(10,0)};
\draw[black,thick,-stealth](11,-3){--++(0,7.35)};

\draw[ultra thick,red,-](12,0) node[above]{\color{black}{A}}--(12+7/1.8,-2/1.8) node[below]{\color{black}{I}};
\draw[black](11,0)--(20.5,0) node[above]{$x$};
\draw[black](11,0)--(11,4.2) node[right]{$t$};
\draw[thick,red,-](12+7/1.8,-2/1.8)--(19,-2) node[above]{\color{black}{B}};
\draw[red,thick,-stealth](19,-2){--++(-1,2/7)};
\draw[red,thick,-stealth](12+7/1.8,-2/1.8){--++(-1,2/7)};
\fill (12+7/1.8,-2/1.8) circle (2.5pt);
\fill (12,0) circle (2.5pt);
\fill (19,-2) circle (2.5pt);
\fill (11,0) circle (2pt) node[below left]{$O$};

\end{tikzpicture}
}
\vskip 0.3cm
\caption{\small{In the frame $\hat O$ on the left, a photon is emitted at A, deposits some of its energy
at the impurity I, and is captured at B with a reduced energy.
In the boosted frame $O$ on the right, the photon is emitted with {\it negative} energy at B, deposits some
energy at the impurity I, and is captured at A with an increased negative energy.
}}\label{HI}
\end{figure}}

Consider a photon of energy $E$ in the frame $\hx,\hht$ emitted by observer A and captured
by observer $B$, transmitting its energy and momentum from A to B (fig.\ 1).
An observer in an appropriately
boosted frame could see the timing of emisstion and absorption reversed, thus observing a
photon originating from B and traveling to A, but with a {\it negative} energy.
This is a tell-tale sign for the observer that this is a frame effect, and the true direction of propagation
was from A to B. Observers {\it cannot} produce photons of negative energy in their own rest frame.

To probe this further, assume that the boosted observer has placed an absorbing impurity I as a sensor
at some point on the path between A and B: on that point the photon will deposit an amount of energy,
thus inducing it to ``start'' with a higher (less negative) energy at B
to end up with the same negative energy at A. nAgain, this is a tell-tale sign that the B to A propagation
is a frame illusion. However, what if the observer decides to place the impurity {\it after} seeing
the photon emitted from B, thus retroactively changing
its (negative) emission energy, for a clear violation of causality? Well, the observer cannot do that, since the
information about the photon's ``emission'' from B cannot be communicated to the position of the impurity
at a speed faster than the speed of the photon. Propagation speeds for positive energy photons
(forward in time)
are always smaller than corresponding speeds for negative energy ones (backward in time), so the signal
cannot arrive before the negative energy photon travels past the impurity.

Another situation is observers A and B relatively at rest exchanging information at maximal speed.
This can certainly happen at an overall back-and-forth speed higher than 1. In the $\hx,\hht$ frame,
and for observers on the $\hx$ axis at positions $\hx_A =0$ and $\hx_B = L$ (and $y=0$), the transmission times
would be $L/c$ with $c>1$ the propagation speed, for an overall back-and-forth communication
time $T = 2L/c$. For two boosted observers in the $x$ direction, still relatively at rest, the two speeds
would be
\be
c_+ = {c-u \over 1-cu} ~,~~~ c_- = {c+u \over 1+cu}
\ee
with $u$ the boost speed. The total communication time would be
\be
T={L\over c_+} + {L\over c_-} = {2Lc(1-u^2)\over c^2-u^2}
\ee
We see that this time can never be negative, since $|u| <1 <|c|$, so communication backward in time is impossible.
Note that reducing the communication time beyond the critical value
\be
T_c = {2Lc \over c^2 +1}
\ee
for which either $c_+$ or $c_-$ diverges, requires
observers in the split wavefront regime. In this case, signal propagation in one direction will happen backward
in {\it coordinate} time, but information can only be exchanged in positive proper time for the
observers.

\section{Phenomenology}

Experimental observation of superluminal and/or anisotropic light propagation would be an indication that we
may live in a braneworld. The phenomenological implications of the tilted/boosted brane
scenario with one extra compact dimension explored in the present work would depend on the parameters
of the brane, as well as the specific frame on this brane in which we live. It would
be tempting to identify our cosmological frame with the ``default'' frame of section {\bf\ref{Tilbo}},
in which case
(\ref{cdefault}) would give the propagation speed of bulk-propagating lightlike fields. However, in the absence
of any specific model justifying this choice, we should stay with a general frame.

To streamline the situation in the general case, we recast equations (\ref{ctilt}) and (\ref{cboost}) for the propagation speed in the common form
\be
c_\phi = {\alpha \beta \cos\phi + \sqrt{1+ \beta^2 - \alpha^2 \cos^2\phi } \over 1- \alpha^2
\cos^2 \phi} ~,~~~\alpha^2 - \beta^2 <1
\ee
with
\bea
&&\alpha= \sqrt{{1-\gamma^2 \cts \over 1-u^2}}  ~,~~ \beta =u \alpha ~~~~~~~\text{for}~~ 
\gamma| \ct\bb |<1 \cr
&&\alpha= \sqrt{{1-\gamma^2 \cts \over 1-u^{-2}}} ~,~~\beta = u^{-1} \alpha ~~~~\text{for}~~ 
\gamma| \ct\bb |>1
\eea
In the above parametrization, $\alpha>0$ and $\beta$ can be treated as phenomenological parameters which,
along with the
anisotropy direction of the coordinate $x$, determine the angular variation of the propagation speed.
The split wavefront regime corresponds to $\alpha>1$, while the tiltlike and boostlike cases correspond to
$|\beta|\bb<\bb\alpha$ and $|\beta|\bb>\bb\alpha$ respectively.
The result (\ref{cdefault}) in the default frame, in particular, is reproduced by $\alpha= \st$
and $\beta=\gamma v\ct$, while the driftless frame results are reproduced by $\beta=0$ for the
tiltlike case and $\alpha= 0$ for the boostlike case. 

The phenomenologically relevant regime is clearly $\alpha \ll1, |\beta|\ll 1$, otherwise macroscopically large deviations
from the speed of light would have been observed. In this regime, the expression for $c_\phi$
approximates to
\be
c_\phi \simeq 1 + \mbox{\large $1\over 2$} (\alpha \cos \phi + \beta)^2
\label{qsmall}\ee
So deviations from the speed of light follow a dipole plus quadrupole angular pattern.
Clearly $c_\phi$ is always superluminal, varying monotonically from $\phi=0$ to $\phi=\pi$ in the
$|\beta|>\alpha$ boostlike case or having a minimum of $c_\phi =1$ at $\cos\phi_o = -\beta/\alpha$ 
in the $|\beta|<\alpha$
tiltlike case. An obserbation of a slight angular variation of the speed of waves or particles consistent with
the above formula would be an indication of the existence of a compact extra dimension.

How we would detect such effects, however, is
another story. Light, for one, is likely to be produced by (and tied to) physics on the brane, and therefore
unable to propagate in the bulk. ``Sterile'' neutrinos may be a candidate for bulk propagation \cite{Die},
and with a speed close to 1 in the bulk they could achieve superluminal velocities on a tilted/boosted brane.
The only waves guaranteed to roam the bulk with abandon are gravitational ones, whose
mere detection is a challenge, let alone the measurement of their exact speed. Nevertheless, a
measurement of their
relative speed compared to photons was performed using a binary neutron star 
event \cite{BPA} and found to deviate from 1 by no more than $\sim 10^{-15}$.
Assuming that photons propagate
on the brane with speed 1, this puts strict bounds to the tilt and boost of any braneworld scenario,
unless we happen to live in a tiltlike braneworld and the relative position of the binary star happened to be
quite close to the direction $\phi_o$, implying a propagation speed close to 1 (see eq. \ref{qsmall}).

\section{Conclusions}

Propagation faster than light seems not just possible but actually generic in braneworlds on compactified
bulk spacetimes, all without tachyons or causality violation. If we live in such a braneworld, and unless
it happens to be in the pristine and quiescent state of zero tilt and zero boost, we should be able to
observe superluminal propagation. In fact, it can be argued that {\it subluminal} propagation would be
problematic: if it existed, we would be able to ``ride'' a wave that propagates at the speed of light
in the bulk, which is impossible no matter what frame or brane we are stuck on.

Apparent superluminal velocities can arise in curved spacetimes, and are always coordinate frame effects.
Co-rotating motion near a Kerr black hole is perhaps the best known example.
In particular, the split wavefront domain corresponds to motion inside the Kerr ergosphere,
where rotation is possible in only one direction and photons can propagate with two different angular
velocities, one of them in negative asymptotic time.

This raises the question of whether there
is an analog of superradiance for waves, or the closely related Penrose process \cite{Penr} for particles
in the braneworld case. In the Kerr
metric, energy can be extracted from the black hole by sending a particle in the
ergosphere and having it decompose into two particles, one of which has negative energy with respect to
the asymptotic region (although a positive energy in its local frame) and subsequently falls into the
black hole, while the other particle escapes and returns to the asymptotic region with increased
energy. This would be a fascinating effect but it does not appear to be realized in the braneworld
scenario considered in this work. This is due to the fact that there is no black hole or other structure
that could absorb negative energy particles and shed energy to the rest of space, and there are no
distinct regions
on the brane, analogous to the ergosphere and asymptotic space in the Kerr metric, with respect to which
a photon would
have different energies. All the energy produced in the brane stays in the brane and has the same value.
{\begin{figure}\vskip-0.5cm
\centering 
{
\begin{tikzpicture}[scale=0.7]


\tikzset{big arrow/.style={decoration={markings,mark=at position 1 with {\arrow[scale=3,#1,>=stealth]{>}}},postaction={decorate},},big arrow/.default=black}

\draw[black,thick,-stealth](-1,0){--++(7,0)};
\draw[black,thick,-stealth](0,-2){--++(0,7)};
\draw[blue, ultra thick,-] (3,0.5)--(3,3);
\draw[green, very thick,-] (3,3)--(1,4);
\draw[green, very thick,-] (3,3)--(5,4);
\draw[blue,very thick,-stealth](3,1){--++(0,1)};
\draw[green,very thick,-stealth](2,3.5){--++(-1,0.5)};
\draw[green,very thick,-stealth](4,3.5){--++(1,0.5)};
\fill (3,3) circle (2.5pt);
\fill (0,0) circle (2pt) node[below left]{$\hat O$};
\draw[black,thick,-stealth](9.2,0){--++(7,0)};
\draw[black,thick,-stealth](10,-2){--++(0,7)};

\draw[blue, ultra thick,-] (11.9,0.3)--(13,2.5);
\draw[green, thick,-] (13,2.5)--(11.5,4.5);
\draw[red, thick,-] (15.5,1.5)--(13,2.5);
\draw[blue,very thick,-stealth](12,0.5){--++(0.5,1)};
\draw[green,thick,-stealth](13,2.5){--++(-1.5,2)};
\draw[red,thick,-stealth](15.5,1.5){--++(-1.25,0.5)};
\fill (13,2.5) circle (2.5pt);

\fill (10,0) circle (2pt) node[below left]{$O$};

\end{tikzpicture}
}
\vskip 0.3cm
\caption{\small{A particle decays into two equal energy superluminal photons in frame $\hat O$ on the left. From the perspective of the boosted frame $O$ on the right, a negative energy photon hits the particle
producing a photon of smaller energy.
}}\label{ER}
\end{figure}
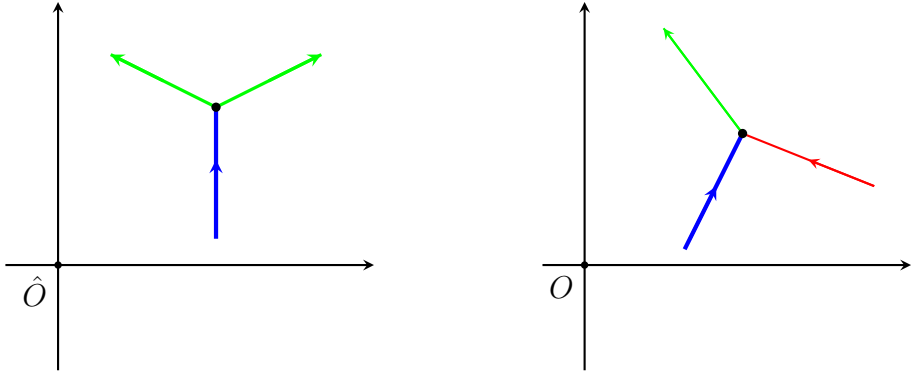}

This still leaves the possibility that, in some process, both positive and negative energy particles are produced
(as measured by a boosted frame), and letting the negative energy ones escape to infinity we are left
locally with increased energy. This scenario is frustrated by the fact that negative energy particles
propagate backwards in time and thus cannot contribute to the energy of future states. As an example,
consider a massive particle decaying into two photons (fig.\ \ref{ER}). In the frame $\hx, \hht$ the photons would
both have positive energy and superluminal velocities. In a boosted frame, one of the photons could
travel backward in time, but then the other one would have an energy necessarily smaller than the energy
of the decaying particle. The photon traveling backward in time would appear as a negative energy
photon colliding with the particle and {\it reducing} its energy while transforming it into another photon.
Although we have not considered all possible processes, such as, e.g., processes where some particles
escape in the bulk and are never recaptured, so far no energy generation occurs.

A scenario in which a Penrose-like effect could occur is one in which the periodic images of the brane
in the bulk have different bulk velocities, and therefore nonzero relative velocities. If the
relative velocity between brane images is negative, a particle emitted by the brane into the bulk
would be recaptured by the
brane with a {\it higher} normal velocity, and therefore a higher energy. In effect, the particle has
extracted energy from the kinetic energy of the brane in the bulk. This, however, would require a
contracting bulk space with a nontrivial spacetime geometry and is not realized in the
locally Minkowski bulk geometry considered in this work. In the absence of bulk contraction it is
not possible for the brane to shed any of its spin energy.

An example closer to the present flat spacetime situation is the spacetime around a spinning cosmic string
\cite{HP}. The energy per unit length of the string will cause the space around the string to develop a deficit
angle, and its angular momentum will cause a time delay/advance for a particle revolving
around the string of magnitude
\be
\tau = 8\pi G J
\ee
with $J$ the angular momentum per unit length of the string. Particles winding around the string in the direction
of its angular momentum will complete a full rotation in a time $t = L/v -\tau$, where $L$ is the proper length
traveled by the particle and $v$ its speed. The effective speed of the particle is
\be
c = {v \over 1-\tau v /L}
\ee
and for $v > L/(L+\tau)$ the effective speed becomes superluminal. Again, no causality violation occurs as long
as $L > \tau$. For $L <\tau$ causality violation {\it would} occur, as in this case there would be closed timelike
curves. However, for this to happen the string should be thinner than $\sim \tau$, so that closed paths of length
$L <\tau$ that enclose its full angular momentum exist. A simple analysis shows that the stress-energy tensor of
strings of such small size would need to be tachyonic in order to produce that much angular momentum.
Therefore, in the absence of tachyonic matter, causality violation is prevented by a kind of cosmic censorship.

By contrast, superluminal propagation on tilted and boosted branes evades any cosmic censorship,
and also does not violate causality.
It therefore provides a window into physics
and phenomena hitherto dismissed as impossible or unphysical. The observation of such phenomena,
of course, will be a challenge.
Still, the ingenuity and imagination of astrophysicists in devising potential tests
of superluminal and/or anisotropy effects should not be underestimated.
At any rate, the theoretical existence of superluminal effects
leaves open an exciting (if wishful) possibility for a glimpse into other dimensions.

\vskip 0.4cm
\noindent
{\bf Acknowledgements}

\noindent

I am thankful to Dan Kabat for discussions that helped me realize the
waveform splitting effect and the correct communication time result, and the anonymous reviewer for
prompting me to consider superradiance effects. This work was supported in part by NSF under grant NSF-PHY-2112729 and by PSC-CUNY under grants 65109-0053 and 6D136-0003.

\end{document}